\documentclass[12pt]{article}
\usepackage{graphicx}
\usepackage{graphics}
\title{Sums of alternating products of Riemann zeta values and solution of a Monthly problem} 
\author{Mark W. Coffey\\
Department of Physics\\
Colorado School of Mines\\
Golden, CO  80401\\
(Received $\mbox{~~~~~~~~~~~~~~~~~~~~~~~~~~~~~~~2011}$)}
\date{July 18, 2011}
\begin{document}
\maketitle
\baselineskip=25 pt
\begin{abstract}

We solve problem 11585 proposed by B. Burdick, AMM June-July 2011 {\bf 118} (6),
p. 558 for the sum of certain products of Riemann zeta function values.  We further
point out an alternating sum analog, and then present and prove different alternating
sum analogs.  In addition, we present summation by parts and other results for the   
Hurwitz and Riemann zeta functions and for the digamma and trigamma functions.

\end{abstract}
 
\medskip
\baselineskip=15pt
\centerline{\bf Key words and phrases}
\medskip 

\noindent

first Stieltjes constant, Euler constant, Riemann zeta function, digamma function,
harmonic number, partial summation 

\vfill
\centerline{\bf 2010 AMS codes} 
11M06, 11Y60

\baselineskip=25pt
\pagebreak
\medskip
\centerline{\bf Solution of problem 11585}
\medskip

B. Burdick has proposed the following problem in the Amer. Math. Monthly {\bf 118}, 
558 (2011).  After giving our solution of this problem, we present several
allied research results.  While an elementary approach to this particular problem
is possible, we follow an approach that uses some special function theory, and it
enables extensions to other problems.  In a later section, we collect summation by
parts expressions for the Hurwitz and Riemann zeta and other functions of interest for use in analytic number theory.  Lastly, we discuss a summatory function of Nielsen \cite{kanemitsu,nielsen}.

To show:
$$\sum_{k=3}^\infty {1 \over k}\left[\sum_{m=1}^{k-2} \zeta(k-m)\zeta(m+1)-k\right]
=3+\gamma^2+2\gamma_1-{\pi^2 \over 3}, \eqno(1.1)$$
where $\zeta$ is the Riemann zeta function, $\gamma=-\psi(1)$ is the Euler constant,
$\psi=\Gamma'/\Gamma$ is the digamma function, $\Gamma$ is the Gamma function, and $\gamma_1$ is the first Stieltjes constant (e.g., \cite{coffeyjmaa}).

{\it Proof}.  One may show that for integers $k \geq 2$
$$2\sum_{n=1}^\infty {{\psi(n)} \over n^k}=k\zeta(k+1)-2\gamma \zeta(k)-\sum_{\ell=1}^
{k-2} \zeta(\ell+1)\zeta(k-\ell),  \eqno(1.2)$$
where $\psi(n)=H_{n-1}-\gamma$ with $H_n=\sum_{k=1}^n 1/k$ the $n$th harmonic number.

We then have
$$\sum_{k=3}^\infty {1 \over k}\left[\sum_{m=1}^{k-2} \zeta(k-m)\zeta(m+1)-k\right]
=\sum_{k=3}^\infty {1 \over k}\left[k\zeta(k+1)-2\gamma \zeta(k)-2\sum_{n=1}^\infty
{{\psi(n)} \over n^k} -k\right].  \eqno(1.3)$$
We decompose the sum on the right side of (1.3) into several others, including first
the elementary ones
$$\sum_{k=3}^\infty {{[\zeta(k)-1]} \over k}=\sum_{k=3}^\infty {1 \over k}
\sum_{r=2}^\infty {1 \over r^k}=\sum_{r=2}^\infty \left[\ln\left({r \over {r-1}} \right)-{1 \over {2r^2}}-{1 \over r}\right]
={3 \over 2} -\gamma-{1 \over 2}\zeta(2), \eqno(1.4)$$
and
$$\sum_{k=3}^\infty [\zeta(k+1)-1]=\sum_{k=4}^\infty \sum_{r=2}^\infty {1 \over r^k}
=\sum_{r=2}^\infty {1 \over {(r-1)}}{1 \over r^3}$$
$$=\sum_{r=2}^\infty \left[-{1 \over r^2}-{1 \over r^3}+\left({1 \over {r-1}}-{1 \over r}\right)\right]=3-\zeta(2)-\zeta(3).  \eqno(1.5)$$
We have
$$\sum_{k=3}^\infty {1 \over k}\sum_{n=1}^\infty {{(H_{n-1}-\gamma)} \over n^k}
=\sum_{k=3}^\infty {1 \over k}\left(\sum_{n=2}^\infty {{(H_{n-1}-\gamma)} \over n^k}
-\gamma\right), \eqno(1.6)$$
with
$$\sum_{k=3}^\infty {1 \over k}\sum_{n=2}^\infty {{(H_{n-1}-\gamma)} \over n^k}
=\sum_{n=2}^\infty (H_{n-1}-\gamma)\left[\ln\left({n \over {n-1}}\right)-{1 \over {2n^2}}
-{1 \over n}\right]$$
$$={{\gamma \zeta(2)} \over 2}-{1 \over 2}\zeta(3)-{\gamma \over 2}+\sum_{n=2}^\infty (H_{n-1}-\gamma)\left[\ln\left({n \over {n-1}}\right)-{1 \over n}\right], \eqno(1.7)$$
and again
$$\sum_{n=2}^\infty \left[\ln\left({n \over {n-1}}\right)-{1 \over n}\right]
=1-\gamma. \eqno(1.8)$$
Above, it is justified to interchange absolutely convergent sums.

Thus
$$\sum_{n=2}^\infty (H_{n-1}-\gamma)\left[\ln\left({n \over {n-1}}\right)-{1 \over n}\right]=\sum_{n=2}^\infty H_{n-1}\left[\ln\left({n \over {n-1}}\right)-{1 \over n}\right]-\gamma(1-\gamma). \eqno(1.9)$$
We now write
$$\sum_{n=2}^\infty H_{n-1}\left[\ln\left({n \over {n-1}}\right)-{1 \over n}\right]
=\sum_{n=1}^\infty H_n\left[\ln\left({{n+1} \over n}\right)-{1 \over {n+1}}\right]$$
$$=\sum_{n=1}^\infty H_n\left[\ln\left({{n+1} \over n}\right)-{1 \over n} +\left({1 \over 
n}-{1 \over {n+1}}\right)\right]$$
$$=-{{\zeta(2)} \over 2}-{\gamma^2 \over 2}-\gamma_1+\zeta(2)={{\zeta(2)} \over 2}-{\gamma^2 \over 2}-\gamma_1.  \eqno(1.10)$$
Here we have used the logarithmic harmonic sum of \cite{kanemitsu} (Lemma 2).
By assembling the above partial results we have
$$\sum_{k=3}^\infty {1 \over k}\left[\sum_{m=1}^{k-2} \zeta(k-m)\zeta(m+1)-k\right]
=\sum_{k=3}^\infty \left\{\left[\zeta(k+1)-1\right]+2\gamma {{[1-\zeta(k)]} \over k}
-{2 \over k}\sum_{n=2}^\infty {{(H_{n-1}-\gamma)} \over n^k} \right\}$$
$$=3-\zeta(2)-\zeta(3)+2\gamma\left({{\zeta(2)} \over 2}+\gamma-{3 \over 2}\right)$$
$$-2\left[-{\gamma \over 2}-{{\zeta(3)} \over 2}+{{\gamma \zeta(2)} \over 2}-\gamma(
1-\gamma)+{{\zeta(2)} \over 2}-{\gamma^2 \over 2}-\gamma_1\right]$$
$$=3-2\zeta(2)+\gamma^2+2\gamma_1.  \eqno(1.11)$$

{\it Remarks}.  Let Li$_s$ be the polylogarithm function.  Then (1.2) may be found by using the integral for $k >1$
$$2\int_0^1 {{[\mbox{Li}_k(t)-t\zeta(k)]} \over {t(t-1)}}dt=k\zeta(k+1)-\sum_{\ell=1}^{k-2}
\zeta(\ell+1)\zeta(k-\ell).  \eqno(1.12)$$
Indeed, observing that $\psi(n)=H_n-\gamma-1/n$, it is seen that (1.2) is Williams' (Euler's) formula \cite{wms}.

For an alternating sum analog of (1.1), see Corollary 3 of \cite{coffey2011}. 
The proof of that Corollary proceeds along a different path than the above proof for (1).

\pagebreak
\centerline{\bf Further sums of zeta products}
\medskip

Furthermore, we have the following
\newline{\bf Proposition 1}.  
$$\sum_{r=1}^\infty {1 \over {(2r+1)}}\left[\sum_{\ell=1}^{2r-1} (-1)^{\ell+1} \zeta(\ell+1)\zeta(2r-\ell+1)-2\right]
=\sum_{n=2}^\infty {1 \over n}\ln\left({n \over {n-1}}\right)
+2-\zeta(2)-2\gamma_1-\gamma^2-\ln 2.  \eqno(1.13)$$

{\it Remarks}.  The approximate numerical value of the sum of (1.13) is $0.262903$.
The remaining logarithmic sum on the right side may be manipulated and rewritten in a 
number of ways, and we return to that later.

{\it Proof}.  One may also show that for $k \geq 3$ an odd integer
$$\sum_{\ell=1}^{k-2} (-1)^{\ell+1}\zeta(\ell+1)\zeta(k-\ell)=2\sum_{n=1}^\infty
{{\psi(n)} \over n^k}+2\zeta(k+1)+2\gamma \zeta(k).  \eqno(1.14)$$
We then have for $k=2r+1$
$$\sum_{r=1}^\infty {1 \over {(2r+1)}}\left[\sum_{\ell=1}^{2r-1} (-1)^{\ell+1} \zeta(\ell+1)
\zeta(2r-\ell+1)-2\right]$$
$$=\sum_{r=1}^\infty {1 \over {(2r+1)}}\left[2\zeta(2r+2)-2+ 2\sum
_{n=1}^\infty {{\psi(n)} \over n^{2r+1}}+2\gamma \zeta(2r+1)\right].  \eqno(1.15)$$
Here
$$\sum_{r=1}^\infty {1 \over {(2r+1)}}\sum_{n=1}^\infty {{(H_{n-1}-\gamma)} \over n^{2r+1}}
=\sum_{r=1}^\infty {1 \over {2r+1}}\left(\sum_{n=2}^\infty {{(H_{n-1}-\gamma)} \over n^{2r+1}}-\gamma\right), \eqno(1.16)$$
$$\sum_{r=1}^\infty {{[\zeta(2r+1)-1]} \over {(2r+1)}}=\sum_{r=1}^\infty {1 \over {(2r+1)}}
\sum_{\ell=2}^\infty {1 \over \ell^{2r+1}}$$
$$=\sum_{\ell=2}^\infty \left[{1 \over 2}\ln\left({{\ell+1} \over {\ell-1}}\right)-{1 \over
\ell}\right]=1-\gamma-{1 \over 2}\ln 2, \eqno(1.17)$$
and
$$\sum_{r=1}^\infty {{[\zeta(2r+2)-1]} \over {(2r+1)}}=\sum_{r=1}^\infty {1 \over {2r+1}}
\sum_{\ell=2}^\infty {1 \over \ell^{2r+2}}=\sum_{\ell=2}^\infty {1 \over \ell}
\left[\mbox{coth}^{-1} \ell-{1 \over \ell}\right]$$
$$=\sum_{\ell=2}^\infty {1 \over \ell}\left[{1 \over 2}\ln\left({{\ell+1} \over {\ell-1}}\right)-{1 \over \ell}\right]={1 \over 2}\sum_{\ell=2}^\infty {1 \over \ell}
\ln\left({{\ell+1} \over {\ell-1}}\right)-\zeta(2)+1.  \eqno(1.18)$$
For the double sum in (1.16) we have
$$\sum_{n=2}^\infty (H_{n-1}-\gamma)\left[{1 \over 2}\ln\left({{n+1} \over {n-1}}\right)
-{1 \over n}\right]=\sum_{n=2}^\infty H_{n-1}\left[{1 \over 2}\ln\left({{n+1} \over {n-1}}\right)-{1 \over n}\right]-\gamma\left(1-\gamma-{1 \over 2}\ln 2\right), \eqno(1.19)$$
and we will show that
$$\sum_{n=2}^\infty H_{n-1}\left[{1 \over 2}\ln\left({{n+1} \over {n-1}}\right)-{1 \over n}\right]={{\zeta(2)} \over 2}-\gamma_1-{\gamma^2 \over 2}-{1 \over 2}\sum_{n=2}^\infty
{1 \over n}\ln \left({{n+1} \over n}\right)-{{\ln 2} \over 2}.  \eqno(1.20)$$
We have
$$\sum_{n=2}^\infty H_{n-1}\left[{1 \over 2}\ln\left({{n+1} \over {n-1}}\right)-{1 \over n}\right]={1 \over 2}\sum_{n=1}^\infty H_n\left[\ln\left({{n+2} \over {n+1}}\right)
-{1 \over {(n+1)}}+\ln\left({{n+1} \over n}\right)-{1 \over {(n+1)}}\right].  \eqno(1.21)$$
The second set of terms on the right side evaluates as in (1.10).  For the rest we again
use \cite{kanemitsu} (Lemma 2) and find
$${1 \over 2}\sum_{n=1}^\infty \left(H_{n+1}-{1 \over {n+1}}\right)\left[\ln\left({{n+2} \over {n+1}}\right)-{1 \over {(n+1)}}\right]
={1 \over 2}\left[{{\zeta(2)} \over 2}-\ln 2-\gamma_1-{\gamma^2 \over 2}-\sum_{n=2}^\infty
{1 \over n}\ln\left({{n+1} \over n}\right)\right].  \eqno(1.22)$$

We have from (1.15) and (1.19) that
$$\sum_{r=1}^\infty {1 \over {(2r+1)}}\left[\sum_{\ell=1}^{2r-1} (-1)^{\ell+1} \zeta(\ell+1)
\zeta(2r-\ell+1)-2\right]=2\sum_{r=1}^\infty {{[\zeta(2r+2)-1]} \over {2r+1}}$$
$$+2\gamma \sum_{r=1}^\infty {{[\zeta(2r+1)-1]} \over {2r+1}}+2\sum_{n=1}^\infty
H_{n-1}\left[{1 \over 2}\ln\left({{n+1} \over {n-1}}\right)-{1 \over n}\right]
-2\gamma \left(1-\gamma-{1 \over 2}\ln 2\right).  \eqno(1.23)$$
Then the Proposition follows by employing the sums of (1.18), (1.17), and (1.20).

{\it Remarks}.  The sum (1.17) is the $t=1$ special case of
$$\sum_{r=1}^\infty {{[\zeta(2r+1)-1]} \over {(2r+1)}}t^{2k+1}={1 \over 2}[\ln \Gamma(2-t)
-\ln \Gamma(2+t)]+(1-\gamma)t, ~~~~~~|t|<2.  \eqno(1.24)$$

For the sum of (1.18) we have the relations
$$2\sum_{k=0}^\infty {{[\zeta(2k+2)-1]} \over {2k+1}}=2\sum_{k=1}^\infty {{\zeta(2k)-1)}
\over {2k-1}}=\sum_{\ell=2}^\infty {1 \over \ell}\ln\left({{\ell+1} \over {\ell-1}}\right)$$
$$=\int_0^1 {{[\psi(2+t)-\psi(2-t)]} \over t}dt.  \eqno(1.25)$$ 
For the sum of the right side of (1.20) we have the integral representation
$$\sum_{n=2}^\infty {1 \over n}\ln\left({{n+1} \over n}\right)
=\sum_{k=1}^\infty {{(-1)^{k+1}} \over k}[\zeta(k+1)-1]$$
$$=\int_0^\infty [\gamma+\Gamma(0,t)+\ln t]{{dt} \over {e^t-1}}-\ln 2$$
$$=\int_0^1 {{(t-1)} \over {\ln
t}}\left[-1-{{\ln(1-t)} \over t}\right]dt, \eqno(1.26)$$
that follows from
$$\ln\left({{n+1}\over n}\right)=\int_0^1 {{t^{n-1}(t-1)} \over {\ln t}}dt.  \eqno(1.27)$$
Here, $\Gamma(x,y)$ is the incomplete Gamma function.  It appears that this sum should be
expressible as $\gamma$ minus a small summatory correction.

The identities of (1.1) and (1.13) can be extended in a number of ways.  For instance, we could include a parameter $t$ so that
$$\sum_{k=3}^\infty {t^k \over k}\left[\sum_{m=1}^{k-2} \zeta(k-m)\zeta(m+1)-k\right]
=\sum_{k=3}^\infty {t^k \over k}\left[k(\zeta(k+1)-1)-2\gamma \zeta(k)-2\sum_{n=1}^\infty
{{\psi(n)} \over n^k}\right].  \eqno(1.28)$$
Here we consider the $t=-1$ case and have
{\newline \bf Proposition 2}. 
$$\sum_{k=3}^\infty {{(-1)^k} \over k}\left[\sum_{m=1}^{k-2} \zeta(k-m)\zeta(m+1)-k\right]
=2\zeta(2)-{1 \over 2}-2\gamma_1-\gamma^2-2\sum_{n=1}^\infty {1 \over n}\ln\left({{n+1}
\over n}\right).  \eqno(1.29)$$

{\it Proof}.  We have from (2)
$$\sum_{k=3}^\infty {{(-1)^k} \over k}\left[\sum_{m=1}^{k-2} \zeta(k-m)\zeta(m+1)-k\right]
=\sum_{k=3}^\infty {{(-1)^k} \over k}\left\{k[\zeta(k+1)-1]-2\gamma \zeta(k) -2\sum_{n=1}^\infty {{\psi(n)} \over n^k}\right\}.  \eqno(1.30)$$
We use
$$\sum_{k=3}^\infty (-1)^k[\zeta(k+1)-1]=\zeta(2)-{1 \over 2}-\zeta(3), \eqno(1.31)$$
being a special case of
$$\sum_{k=1}^\infty t^k[\zeta(k+1)-1]={t \over {t-1}}-\gamma-\psi(1-t), \eqno(1.32)$$
$$\sum_{k=3}^\infty {{(-1)^k} \over k}[\zeta(k)-1]=\ln 2-{1 \over 2}+\gamma -{{\zeta(2)}
\over 2}, \eqno(1.33)$$
being a special case of
$$\sum_{k=2}^\infty {{(-1)^k} \over k}[\zeta(k)-1]=t(1-\gamma)+\ln(1-t)+\ln \Gamma(1-t),
\eqno(1.34)$$
and
$$\sum_{k=3}^\infty {{(-1)^k} \over k}\sum_{n=1}^\infty {{(H_{n-1}-\gamma)} \over n^k}
=\sum_{k=3}^\infty {{(-1)^k} \over k}\left(\sum_{n=2}^\infty {{(H_{n-1}-\gamma)} \over n^k}
-\gamma\right), \eqno(1.35)$$
with
$$\sum_{k=3}^\infty {{(-1)^k} \over k}\sum_{n=2}^\infty {1 \over n^k}=\sum_{n=2}^\infty
\left[\ln\left({n \over {n+1}}\right)+{1 \over n}-{1 \over {2n^2}}\right]$$
$$=-{1 \over 2} +\gamma-{{\zeta(2)} \over 2}+\ln 2.  \eqno(1.36)$$
We also require the following sum, evaluated with another application of \cite{kanemitsu}
(Lemma 2):
$$\sum_{n=2}^\infty H_{n-1}\left[\ln\left({n \over {n+1}}\right)+{1 \over n}-{1 \over {2n^2}}\right]=\sum_{n=2}^\infty \left(H_n-{1 \over n}\right)\left[-\ln\left({{n+1} \over n}\right)+{1 \over n}-{1 \over {2n^2}}\right]$$
$$={{\zeta(2)} \over 2}+\gamma_1+{\gamma^2 \over 2}-\zeta(3)-\sum_{n=1}^\infty{1 \over n}
\left[-\ln\left({{n+1} \over n}\right)+{1 \over n}-{1 \over {2n^2}}\right]$$
$$=-{{\zeta(2)} \over 2}+\gamma_1+{\gamma^2 \over 2}-{{\zeta(3)} \over 2} +\sum_{n=1}^\infty{1 \over n}\ln\left({{n+1}\over n}\right).  \eqno(1.37)$$
Combining (1.31) with minus twice (1.37) yields the Proposition.

{\it Remark}.  The approximate numerical value of the alternating sum of Proposition 2
is $0.0868281269673$, being roughly the negation of the value of the sum of (1),
$-0.102321900856$.

We next consider an additionally alternating sum and have the following.
\newline{\bf Proposition 3}.
$$\sum_{r=1}^\infty {{(-1)^r} \over {(2r+1)}}\left[\sum_{\ell=1}^{2r-1} (-1)^{\ell+1} \zeta(\ell+1)\zeta(2r-\ell+1)-2\right]$$
$$=2\sum_{\ell=2}^\infty {{\cot^{-1}\ell} \over \ell}-2\zeta(2)+2
+2\sum_{n=2}^\infty H_{n-1}\left(\cot^{-1}n-{1 \over n}\right).  \eqno(1.38)$$

{\it Proof}.  Now from (1.14) we have
$$\sum_{r=1}^\infty {{(-1)^r} \over {(2r+1)}}\left[\sum_{\ell=1}^{2r-1} (-1)^{\ell+1} \zeta(\ell+1)\zeta(2r-\ell+1)-2\right]$$
$$=\sum_{r=1}^\infty {{(-1)^r} \over {(2r+1)}}\left[2[\zeta(2r+2)-1]+ 2\sum
_{n=1}^\infty {{\psi(n)} \over n^{2r+1}}+2\gamma \zeta(2r+1)\right],  \eqno(1.39)$$
and we recall the series for $x^2 \geq 1$
$$\cot^{-1} x = \sum_{k=0}^\infty {{(-1)^k} \over {(2k+1)}}{1 \over x^{2k+1}}.
\eqno(1.40)$$
Here
$$\sum_{r=1}^\infty {{(-1)^r}\over {(2r+1)}}[\zeta(2r+1)-1] =\sum_{r=1}^\infty 
{{(-1)^r} \over {(2r+1)}}\sum_{\ell=2}^\infty {1 \over \ell^{2r+1}}$$
$$=1-\gamma-{\pi \over 4}+{i \over 2}\ln\left({{\Gamma(1+i)} \over {\Gamma(1-i)}}\right), \eqno(1.41)$$
and
$$\sum_{r=1}^\infty {{(-1)^r} \over {(2r+1)}}[\zeta(2r+2)-1]=\sum_{r=1}^\infty 
{{(-1)^r} \over {2r+1}}\sum_{\ell=2}^\infty {1 \over \ell^{2r+2}}=\sum_{\ell=2}^\infty {1 \over \ell}\left[\cot^{-1} \ell-{1 \over \ell}\right]$$
$$=\sum_{\ell=2}^\infty {{\cot^{-1}\ell} \over \ell}-\zeta(2)+1.  \eqno(1.42)$$

Writing
$$\sum_{r=1}^\infty {{(-1)^r} \over {(2r+1)}}\sum_{n=1}^\infty {{(H_{n-1}-\gamma)} \over n^{2r+1}}
=\sum_{r=1}^\infty {{(-1)^r} \over {2r+1}}\left(\sum_{n=2}^\infty {{(H_{n-1}-\gamma)} \over n^{2r+1}}-\gamma\right)$$
$$=\sum_{n=2}^\infty H_{n-1} \left(\cot^{-1}n-{1 \over n}\right)-\gamma\sum_{r=1} ^\infty {{(-1)^r}\over {(2r+1)}}[\zeta(2r+1)-1], \eqno(1.43)$$
we see that the overall contribution of (1.41) in (1.39) is annulled.  Hence the
Proposition follows.

{\it Remarks}.  The approximate numerical value of the sum of (1.38) is $-0.215191890953$.
The harmonic arc cotangent sum on the right side of that equation is subject to further investigation.  After all,
$$\cot^{-1} n-{1 \over n}={1 \over {2i}}\ln\left({{in-1}\over {in+1}}\right)-{1 \over n}={1 \over {2i}}\left[-\ln\left({{in+1} \over {in}}\right)-\ln\left({{in} \over
{in-1}}\right)\right]-{1 \over {2n}}-{1 \over {2n}}.  \eqno(1.44)$$

The sum on the right side of (1.42) is a case of the following integral representation.
{\bf Lemma 1}.  For $|z| \leq 1$,
$$\sum_{\ell=2}^\infty z^\ell {{\cot^{-1}\ell} \over \ell}=-\int_0^1 {{\sin(\ln t)}
\over {t \ln t}}\ln(1-zt)dt-{\pi \over 4}z.  \eqno(1.45)$$

{\bf Corollary 1}.
$$-\int_0^1 {{\sin(\ln t)}\over {t \ln t}}\ln(1-t)dt=\sum_{\ell=1}^\infty H_\ell
[\cot^{-1}\ell-\cot^{-1}(\ell+1)].  \eqno(1.46)$$

{\it Proof}.  First we have for $|a| \leq 1$,
$$\sum_{k=1}^\infty {a^k \over k}\ln\left({{x+k} \over {y+k}}\right)=\sum_{k=1}^\infty
\int_0^1 [t^{x+k-1}-t^{y+k-1}]{{dt} \over {\ln t}}$$
$$=-\int_0^1 {{(t^x-t^y)} \over {t \ln t}}\ln(1-at)dt.  \eqno(1.47)$$

Then
$$\sum_{\ell=2}^\infty z^\ell {{\cot^{-1}\ell} \over \ell}={1 \over {2i}}\sum_{\ell=2}^
\infty {z^\ell \over \ell}\ln\left({{i\ell-1} \over {i\ell+1}}\right)$$
$$={1 \over {2i}}\sum_{\ell=2}^\infty {z^\ell \over \ell}\ln\left({{\ell+i} \over {\ell-i}}\right)$$
$$={1 \over {2i}}\left[-\int_0^1 {{(t^i-t^{-i})} \over {t\ln t}}\ln(1-zt)dt-z\ln\left(
{{1+i} \over {1-i}}\right)\right]$$
$$=-\int_0^1 {{\sin(\ln t)}\over {t \ln t}}\ln(1-zt)dt-{\pi \over 4}z.  \eqno(1.48)$$

We use summation by parts to obtain the Corollary.

(1.47) appears to provide a correction to the integral representation given 
as (44.8.5) in \cite{hansen} (p. 289).

It is a simple matter to show that expansion of the $ln$ factor in the integrand
of the right side of (1.45) returns the original summation.  Related to this, we
provide a quick derivation of the integral
$$I(k) \equiv \int_0^\infty {{\sin u} \over u}e^{-ku} du=\cot^{-1} k.  \eqno(1.49)$$

We have $I(0)=\pi/2$ with
$${{dI(k)} \over {dk}}=-\int_0^\infty \sin ue^{-ku} du=-{1 \over {1+k^2}}. \eqno(1.50)$$
Hence $I(k)=\int_0^1 {{\sin(\ln t)} \over {\ln t}}t^{k-1}dt=\cot^{-1} k$.

\medskip
\centerline{\bf Results from partial summation}
\medskip

We collect in the following expressions resulting from the use of partial 
summation.  In particular, part (c) is a generalization of (2.23) in \cite{kanemitsu}.
As usual, $\zeta(s,a)$ denotes the Hurwitz zeta function and $\psi^{(j)}$ the
polygamma functions, with $\psi'$ the trigamma function.
{\newline \bf Lemma 2}.  (a)
$$\zeta(s+1)=\sum_{r=1}^\infty H_r\left({1 \over r^s}-{1 \over {(r+1)^s}}\right),
\eqno(2.1)$$
(b)
$$(1-2^{-s})\zeta(s+1)=\sum_{r=1}^\infty H_r\left[{{(-1)^{r+1}} \over n^s}-{{(-1)^r}
\over {(n+1)^s}}\right], \eqno(2.2)$$
(c)
$$\zeta(s+1,a)=\sum_{r=0}^\infty [\psi(a+r+1)-\psi(a)]\left[{1 \over {(r+a)^s}}-
{1 \over {(r+a+1)^s}}\right], \eqno(2.3)$$
(d)
$$\psi(x)=-\gamma-{1 \over x}+x\sum_{k=1}^\infty H_k\left[{1 \over {x+k}}-{1 \over {x+k+1}}\right], \eqno(2.4)$$
(e)
$$\psi'(x)={1 \over x^2}+\sum_{k=1}^\infty H_k\left[{1 \over {(x+k)}}-{1 \over {(x+k+1)}}\right]+x\sum_{k=1}^\infty H_k\left[{1 \over {(x+k+1)^2}}-{1 \over {(x+k)^2}}\right], \eqno(2.5)$$
and (f)
$$\psi^{(j)}(x)=(-1)^{j+1}j!\left \{{1 \over x^{j+1}}+\sum_{k=1}^\infty H_k\left[{1 \over {(x+k)^j}}-{1 \over {(x+k+1)^j}}\right] \right.$$
$$\left.+x\sum_{k=1}^\infty H_k\left[{1 \over {(x+k+1)^{j+1}}}-{1 \over {(x+k)^{j+1}}}\right] \right\}. \eqno(2.6)$$

{\it Proof}.  (a) is obviously a special case of (c).  For (c) we write for Re $s>0$ and Re $a>0$
$$\zeta(s+1,a)=\sum_{n=0}^\infty {1 \over {(n+a)^{s+1}}}=\sum_{n=0}^\infty {1 \over {(n+a)}}{1 \over {(n+a)^s}}, \eqno(2.7)$$
and apply summation by parts.  Part (b) is based upon the alternating zeta function
for Re $s>0$
$$\sum_{r=1}^\infty {{(-1)^{r+1}} \over n^s}=(1-2^{1-s})\zeta(s).  \eqno(2.8)$$
Part (d) is based upon
$$\psi(x)=-\gamma-{1 \over x}+x\sum_{k=1}^\infty {1 \over {k(x+k)}}, \eqno(2.9)$$
and (e) and (f) follow by repeated differentiation.


We have
$$\lim_{s\to 0}[\zeta'(s+1,a)-\zeta'(s+1,b)]=\gamma_1(b)-\gamma_1(a),  \eqno(2.10)$$
and it must follow that $\gamma_1-\gamma_1(1/2)=\ln^2 2+2\gamma \ln 2$.

We note that Lemma 2 may also be applied to expressions such as
$$\lim_{s \to 0}[\zeta'(s+1,a)+\psi'(s)]=\zeta(2)-\gamma_1(a),  \eqno(2.11)$$
where
$$\zeta'(s+1,a)=\sum_{r=0}^\infty [\psi(a+r+1)-\psi(a)]\left[{{\ln(r+a+1)} \over {(r+a+1)^s}}-{{\ln(r+a)} \over {(r+a)^s}}\right]. \eqno(2.12)$$

\medskip
\centerline{\bf On a function $\xi(x)$}
\medskip

Nielsen \cite{nielsen} introduced a function $\xi(x)$ such that
$$[\psi(x)+\gamma]^2=\psi'(x)-\zeta(2)-2\xi(x).  \eqno(3.1)$$
Explicitly,
$$\xi(x)=\sum_{n=1}^\infty H_n\left({1 \over {x+n}}-{1 \over {n+1}}\right), \eqno(3.2)$$
and we will give integral representations for both $\xi(x)$ and its integral.

From either Lemma 2 of \cite{kanemitsu} or Proposition 3 of \cite{coffey2011}, it follows that
$$\int_0^1 \xi(x)dx=\sum_{n=1}^\infty H_n \left[\ln\left({{n+1} \over n}\right)
-{1 \over {n+1}}\right]=\sum_{n=1}^\infty H_n \left[\ln\left({{n+1} \over n}\right)
-{1 \over n}\right]+\zeta(2)$$
$$={1 \over 2}[\zeta(2)-\gamma^2-2\gamma_1].  \eqno(3.3)$$

{\bf Lemma 3}.  For $x$ not a negative integer, 
$$\xi(x)=\int_0^\infty [e^{(1-x)t}-1]{{\ln(1-e^{-t})} \over {(1-e^t)}}dt$$
$$=\int_0^1 (u^{x-1}-1){{\ln(1-u)} \over {u-1}}du, \eqno(3.4)$$
and
$$\int_0^1 \xi(x)dx={1 \over 2}[\zeta(2)-\gamma^2-2\gamma_1]
=-\int_0^\infty \left[{1 \over t}+{1 \over {1-e^t}}\right]\ln(1-e^{-t})dt$$
$$=\int_0^1\left[{1 \over {u\ln u}}-{1 \over {u-1}}\right]\ln(1-u)du.  \eqno(3.5)$$

{\it Proof}.  We write
$$\xi(x)=\sum_{n=1}^\infty H_n\left(\int_0^\infty e^{-(x+n)t}dt-\int_0^\infty e^{-(n+1)t}dt\right).  \eqno(3.6)$$
Justified by absolute convergence we may interchange the summation and integration,
use the generating function for harmonic numbers, and (3.4) follows.  For (3.5) we carry out the indicated integration from (3.4).

{\it Remark}.  From (3.4) we recover the value
$$\xi(0)=-\int_0^\infty \ln(1-e^{-t})dt=-\int_0^1 {{\ln(1-u)} \over u}du=\zeta(2).  \eqno(3.7)$$


\pagebreak

\end{document}